\documentstyle[aps,epsf]{revtex}
\begin{document}
\draft
\title{The nematic-isotropic phase transition in linear fused hard-sphere
chain fluids}
\author{K.M.~Jaffer, S.B.~Opps and D.E.~Sullivan}
\address{Department of Physics and Guelph-Waterloo Program for Graduate
Work in Physics\\ University of Guelph, Guelph, Ontario N1G 2W1, Canada}
\date{\today}
\maketitle

\begin{abstract}
 We present a modification of the generalized Flory dimer theory to
investigate the nematic (N) to isotropic (I) phase transition in chain
fluids.  We focus on rigid linear fused hard-sphere (LFHS) chain molecules
in this study.  A generalized density functional theory is developed,
which involves an angular weighting of the dimer reference fluid as
suggested by decoupling theory, to accommodate nematic ordering in the
system.  A key ingredient of this theory is the calculation of the exact
excluded volume for a pair of molecules in an arbitrary relative
orientation, which extends the recent work by Williamson and Jackson for
linear tangent hard-sphere chain molecules to the case of linear fused
hard-sphere chains with arbitrary intramolecular bondlength. The present
results for the N-I transition are compared with previous theories and
with computer simulations.  In comparison with previous studies, the
results show much better agreement with simulations for both the
coexistence densities and the nematic order parameter at the transition. 
\end{abstract}
\vspace{1cm}

\section{Introduction}	\label{sec:intro}

 The hard-sphere chain fluid has been widely studied in recent years as a
model for polymeric and related liquids.  Various theories have been
applied to determine the thermodynamic properties of the model fluid, such
as thermodynamic perturbation theory (TPT)\cite{1,2a,2b,3,4}, generalized
Flory dimer theory (GFD)\cite{5,6,7}, scaled particle theory
(SPT)\cite{bn,8,17b} and various integral-equation
theories\cite{10a,10b,11,12}.  Despite these studies, several limitations
still characterize most theories for the hard-sphere chain fluid,
including the independence of intramolecular structure on the surrounding
medium in homogeneous fluids\cite{2a,2b}, the inability to account for
liquid-crystalline ordering\cite{2a,2b,7,12} and a general difficulty in
formulating practical extensions of these theories to non-uniform systems.
Several density-functional approaches\cite{13,14a,14b} to the latter
problem have been examined, but these have not overcome the other
limitations mentioned above. 

In this article, we investigate a systematic and straightforward
density-functional method to begin resolving these limitations. We focus
on a rather specific problem, namely to characterize the nematic-isotropic
(N-I) phase transition in a fluid composed of {\it rigid} linear fused
hard-sphere (LFHS) chain molecules.  Several related works have examined
this topic recently\cite{7,17b,12,15,16,17a}.  Mehta and Honnell\cite{7}
applied GFD theory to the LFHS fluid, but were restricted to examining the
isotropic phase.  This is also true of the integral-equation and SPT
approaches studied in ref.\cite{12}, although some information about the
N-I transition could be deduced from stability analysis of the isotropic
phase.  Refs.\cite{17b,15,16,17a} studied the N-I transition in linear
tangent hard-sphere (LTHS) chain fluids, using a density-functional
approach based on the decoupling approximation introduced by
Parsons\cite{18} and Lee\cite{19} for hard ellipsoids and hard
spherocylinders.  This approximation consists of a rescaling of the virial
series originally used by Onsager\cite{20} to account for the
thermodynamic properties of a rigid-rod fluid. (Ref.\cite{17a} also
describes an extension of the method to a fluid of semi-flexible chains.)

The formulation discussed here is very similar to that of
refs.\cite{15,16,17a}, but utilizes a concept closely related to that of
the TPT and GFD theories, namely that the thermodynamic properties of the
full chain fluid can be described by a judicious combination of those
characterizing ``reference'' fluids composed of smaller sub-units, in
particular monomers and dimers.  Both formally and intuitively, this
picture is well suited to describing fluids of {\it flexible} chain
molecules\cite{1,2a,2b,3,4,5,6}.  A key ingredient of the present work,
focusing on rigid rods, is that the excluded volume (and hence the second
virial coefficient) for a pair of LFHS molecules in an arbitrary
orientation can be written {\it exactly} as a linear combination of that
for a pair of diatomic hard-sphere molecules and a pair of hard-sphere
monomers.  This property is based on a recent analysis of the
orientation-dependent excluded volume for LTHS chains by Williamson and
Jackson\cite{21}.  Here it is shown that the results of ref.\cite{21} can
be extended straightforwardly to linear {\it fused} hard-sphere chains of
arbitrary intramolecular bondlength.

In the present article, the decoupling approximation is utilized in two
stages.  In the first stage, it is assumed that all higher-order virial
coefficients scale (in terms of monomer and diatomic contributions) in the
same way as the second virial coefficient.  In the case of an isotropic
fluid phase, the resulting approximation for thermodynamic properties is
similar to that of GFD theory, but differs from the latter in reproducing
the exact second virial coefficient.  In the second stage, the excess free
energy of the reference dimer fluid is rescaled in terms of the free
energy of the isotropic phase of this fluid.  This rescaling incorporates
the exact weighted angle-average of the pair excluded volume, as in the
theories of refs.\cite{15,16,17a}, and thus is able to account for nematic
ordering, in contrast to the original GFD theory.

The theory is presented in Section \ref{sec:theory}. Although the focus of
this paper is on uniform phases of a rigid molecular fluid, Section
\ref{sec:gen} begins with the more general case of a non-uniform fluid and
extensions of the theory to semi-flexible molecules are indicated briefly.
In Sections \ref{sec:ev} and \ref{sec:fe}, the theory is specialized to
uniform phases of LFHS chains.  Calculations and results are presented in
Section \ref{sec:results}, including comparisons with available Monte
Carlo data for the N-I transition of both LTHS and LFHS chains.  It is
found that the present theory overestimates the reduced pressure when
compared with both simulations and previous theories.  However, the
present results for coexistence densities and the nematic order parameter
at the transition are in excellent agreement with simulations and are more
accurate than those predicted by other recent
theories\cite{17b,15,16,17a}.  Section \ref{sec:conclude} contains a
summary and some conclusions.

\section{Theory}	\label{sec:theory}
\subsection{General}    \label{sec:gen}

 Here we consider a one-component fluid containing on average $N$
molecules, each molecule (or ``$n$-mer'') consisting of a rigid linear
chain of $n$ atomic sites.  The configuration of any such molecule
labelled $i$ is specified by the position ${\bf r}_i$ of some arbitrarily
chosen ``center'' of the molecule and by the orientational Euler angles
$(\theta_i,\phi_i) \equiv \omega_i$ of the molecular chain axis relative
to some space-fixed frame.  We assume that the total intermolecular
potential energy for $N$ molecules is pairwise additive, and denote the
pair potential between molecules $i$ and $j$ as $U_{(2)}({\bf
r}_i\omega_i,{\bf r}_j\omega_j)$. 

The probability density for finding any molecule in a configuration ${\bf
r}\omega$ is denoted $\rho({\bf r}\omega)$, which is normalized such that
\begin{equation}	\label{eq:1}
 \int d{\bf r} d\omega~\rho({\bf r}\omega)~ =~ N .
\end{equation}
The integration over molecular position is constrained to occur within the
system volume $V$.  In density-functional theory, the Helmholtz free
energy $F$ of the fluid is expressed as a functional of
$\rho(\bf{r}\omega)$.  A formally exact, generalized virial expansion of
$F$ in powers of $\rho(\bf{r}\omega)$ can be obtained by straightforward
adaptation of that for monatomic fluids \cite{22}, and is given by
\begin{eqnarray}	\label{eq:2}
 F & = & kT \int d{\bf r} d\omega \rho({\bf r}\omega) \lbrack
\ln(4\pi\rho({\bf r}\omega)\Lambda)-1 + \beta
U_{ex}({\bf r}\omega) \rbrack + \Delta F ,
\end{eqnarray}
where $T$ is temperature, $k$ is Boltzmann's constant, $\beta=(kT)^{-1}$,
$\Lambda$ is the thermal de Broglie ``volume'' of a molecule, and $U_{ex}$
is any external potential field acting on the fluid.  $\Delta F$ is the
excess (over an ideal gas) Helmholtz free energy of the system, and is
expressed by the virial series
\begin{equation}	\label{eq:3}
 \Delta F = kT \sum_{m=2}^{\infty} \frac{\hat{B}_m}{m-1} .
\end{equation}
The generalized virial coefficients $\hat{B}_{m}$ have a standard
diagrammatic representation in terms of irreducible cluster integrals,
where the vertices of the diagrams represent the product of functions
$\rho({\bf r}_1\omega_1),\rho({\bf r}_2\omega_2),\ldots \rho({\bf
r}_m\omega_m)$ and the bonds denote the Mayer function
\begin{equation}	\label{eq:4}
 f({\bf i},{\bf j}) = e^{-\beta 
U_{(2)}({\bf i},{\bf j})} - 1 .
\end{equation}
For simplicity, the label $\mathbf{i}$ is used to denote ${\bf
r}_i\omega_i$.  Thus,
\begin{eqnarray}	\label{eq:5a}
 \hat{B}_{2} & = & -\frac{1}{2} \int d{\bf 1}d{\bf 2}
\rho({\bf 1}) \rho({\bf 2}) f({\bf 1},{\bf 2})
,\\	\label{eq:5b}
\hat{B}_{3} & = & -\frac{1}{3} \int d{\bf 1}d{\bf 2}
d{\bf 3}\rho({\bf 1}) \rho({\bf 2})
\rho({\bf 3}) f({\bf 1},{\bf 2}) 
f({\bf 2},{\bf 3}) f({\bf 3},{\bf 1}) ,
\end{eqnarray}
etc.

The preceding equations can be formally generalized to fluids of
semi-flexible molecules on interpreting the symbol $\mathbf{i}$ to include
all internal coordinates of molecule $i$ and replacing
$U_{ex}(\mathbf{i})$ by the full one-body potential incorporating internal
bonding and flexibility constraints \cite{23a,23b}.  In the usual approach
of density-functional theory, the distribution function $\rho(\mathbf{i})$
in the preceding expressions for $F$ is considered to be an arbitrary
function of the molecular configuration $\mathbf{i}$, while the
equilibrium form of that distribution function is obtained by functional
minimization of the corresponding grand canonical potential $\Omega = F -
N\mu$, where $\mu$ is the chemical potential.

In the standard application of the decoupling approximation for the
case of {\it uniform} fluids\cite{18,19,30}, the $2$nd virial coefficient
is treated exactly while the $m$th-order virial coefficient is
approximated as
\begin{equation}	\label{eq:6}
\hat{B}_{m} = \frac{\hat{B}_{2}}{\hat{B}_{2}^{ref}}\hat{B}_{m}^{ref}~,
\end{equation}
where $\hat{B}_{m}^{ref}$ is the virial coefficient of an appropriate
reference fluid. In previous works, the reference fluid has been chosen to
be either a hard-sphere fluid \cite{18,19} or the {\it isotropic} phase of
the actual system being considered \cite{15,16,30}. In either case,
resummation of the virial series (\ref{eq:3}) then gives
\begin{equation}	\label{eq:7}
\Delta F =  \frac{\hat{B}_{2}}{\hat{B}_{2}^{ref}}\Delta F^{ref}. 
\end{equation}
(A generalization of this procedure for a non-uniform fluid of hard
spherocylinders is described in ref.\cite{31}.)  The GFD and TPT theories
for fluids of chain molecules are based on a different choice of
reference system, involving the decomposition of each molecule
into its constituent monomers and constituent dimers. Although originally
derived using quite different arguments, we can obtain the GFD theory
using the present framework. First, at the level of the $2$nd virial
coefficient, we postulate that $\hat{B}_{2}$ can be written as
\begin{equation}	\label{eq:8}
\hat{B}_{2} = a_{n}^{(M)}\hat{B}_{2}^{(M)} +
a_{n}^{(D)}\hat{B}_{2}^{(D)}~,
\end{equation}
where $a_n^{(M)}$ and $a_n^{(D)}$ are appropriate mixing parameters.
Here, $\hat{B}_{2}^{(M)}$ is the generalized $2$nd virial coefficient
characterizing the fluid which results on decomposing each chain molecule
into its constituent monomers, while $\hat{B}_{2}^{(D)}$ is the analogous
quantity for the reference fluid of constituent dimers.  Formally, the
latter decomposition assumes that the number of monomers per chain $n$ is
even. However, we anticipate (as suggested by previous work on GFD and
related theories \cite{4,6,26}) that both the monomer and dimer reference
fluids should be characterized by suitable {\it effective} values of the
number of monomers $n_{M}$ and number of dimers $n_{D}$, respectively.
In accordance with the decoupling approximation, we assume that the
mixing parameters $a_{n}^{(M)}$ and $a_{n}^{(D)}$ can be chosen so that
(\ref{eq:8}) generates the exact $\hat{B}_{2}$. This condition will be
examined in subsection \ref{sec:ev}. If it is assumed that all
higher-order virial coefficients $\hat{B}_{m}$ scale in a similar manner,
$i.e.$
\begin{equation}	\label{eq:9}
\hat{B}_{m} = a_{n}^{(M)}\hat{B}_{m}^{(M)} +
a_{n}^{(D)}\hat{B}_{m}^{(D)}~,
\end{equation}
then resummation of the series in (\ref{eq:3}) gives
\begin{equation}	\label{eq:10}
 \Delta F = a_{n}^{(M)} \Delta F^{(M)}[\rho_{M}({\bf r})] + a_{n}^{(D)} \Delta 
F^{(D)}[\rho_{D}({\bf r}\omega)] .
\end{equation}
In this equation, $\rho_M({\bf r})$ and $\rho_D({\bf r}\omega)$ are the
probability densities for monomers and dimers, respectively, while $\Delta
F^{(M)}$ and $\Delta F^{(D)}$ are the corresponding excess free energies,
which in general are functionals of the densities. For simplicity, we
assume that the $n$-mer contains only one type of monomer and dimer
sub-unit; the generalization to heteronuclear molecules is
straightforward. 

The standard GFD theory for isotropic, homonuclear hard-sphere chain
fluids, as investigated previously in refs.\cite{5,6,7}, can be formulated
from the theory presented here.  In particular, GFD theory is obtained
from eqs.(\ref{eq:2}), (\ref{eq:10}), and a {\it modified} form of the
condition (\ref{eq:8}).  In order to determine the parameters $a_n^{(M)}$
and $a_n^{(D)}$ in this case, (\ref{eq:8}) is applied with $\hat{B}_2$ and
$\hat{B}_2^{(D)}$ {\it approximated} by the cross second virial
coefficient between a monomer and the full $n$-mer and between a monomer
and a dimer, respectively.  An additional condition is then required to
fix $a_n^{(M)}$ and $a_n^{(D)}$ uniquely, which turns out to be
\begin{equation}	\label{eq:11}
 n_M a_n^{(M)} + n_D a_n^{(D)} = 1 .
\end{equation}
This is consistent with the ``initial conditions'' that
$a_n^{(M)} = 1$, $a_n^{(D)} = 0$ when $n = n_M = 1$ and $a_n^{(M)} = 0$,
$a_n^{(D)} = 1$ when $n = n_M = 2n_D = 2$.  With these approximations, one
obtains
\begin{equation}	\label{eq:12}
 n_M a_n^{(M)} = -Y_n \quad , \quad n_D a_n^{(D)} = 1 + Y_n ,
\end{equation}
where $Y_n$ is the quantity defined in eq.(14) of Honnell and
Hall\cite{5}.

In the next subsection, we shall specialize (\ref{eq:8}) and (\ref{eq:10}) 
to the case of a uniform fluid of rigid homonuclear hard-sphere chains. It
will be shown that (\ref{eq:8}) can be satisfied {\it exactly} when using
the exact values of $\hat{B}_2$ and $\hat{B}_2^{(D)}$ and with a unique
choice for the parameters $a_n^{(M)}$ and $a_n^{(D)}$ depending only on
the geometric properties of the molecules.

\subsection{Application to LFHS Chains}	  \label{sec:ev}

 Here we consider that the full chain molecule is a rigid rod composed of
$n$ identical hard-sphere atoms, each of diameter $d$, with adjacent atoms
separated by bondlength $l$.  In a {\it uniform} but possibly
orientationally ordered fluid, the $n$-mer probability density takes the
form
\begin{equation}	\label{eq:13}
 \rho({\bf r}\omega) = \rho f(\omega) ,
\end{equation}
where $\rho = N/V$ is the molecular number density and $f(\omega)$ is a
normalized angular distribution function.  The generalized second virial
coefficient $\hat{B}_{2}$ defined by (\ref{eq:5a}) becomes
\begin{equation}	\label{eq:14}
 \hat{B}_{2} = -\frac{\rho^{2}V}{2} \int d{\bf r}_{12} d\omega_1
d\omega_2 f(\omega_1) f(\omega_2)
f({\bf r}_{12},\omega_1,\omega_2) .
\end{equation}
For hard-body interactions, this reduces to
\begin{equation}	\label{eq:15}
 \hat{B}_{2} = \frac{\rho^{2}V}{2} \int d\omega_{1} d\omega_{2}
f(\omega_{1}) f(\omega_{2}) v_{e}^{(n)}(\theta_{12}) ,
\end{equation}
where $v_{e}^{(n)}(\theta_{12})$ is the excluded volume between two rigid
$n$-mers, which depends on the relative angle $\theta_{12}$ between their
axes.  Analogous expressions hold for the monomer and dimer virial
coefficients, $\hat{B}_{2}^{(M)}$ and $\hat{B}_{2}^{(D)}$.  In particular,
\begin{equation}	\label{eq:16}
 \hat{B}_{2}^{(D)} = \frac{\rho_{D}^{2}V}{2} \int d\omega_{1} d\omega_{2}
f(\omega_{1}) f(\omega_{2}) v_{e}^{(2)}(\theta_{12}) ,
\end{equation}
where $\rho_{D} = n_{D}\rho$ is the number density of dimers and
$v_{e}^{(2)}(\theta_{12})$ is the corresponding excluded volume.  Note
that for rigid linear rods, the same angular distribution function,
$f(\omega)$, characterizes the full rod and any diatomic subunit of the
rod.  For the monomer fluid,
\begin{equation}	\label{eq:17}
 \hat{B}_{2}^{(M)} = \frac{\rho_{M}^{2}V}{2} v_{e}^{(1)} ,
\end{equation}
where $\rho_M = n_M\rho$ is the monomer number density.  The monomer
excluded volume $v_e^{(1)}$, of course, has no angular dependence. 
Requiring that (\ref{eq:8}) holds exactly then implies the following
relation between the excluded volumes, {\it independent of the angular
distribution function and number densities},
\begin{equation}	\label{eq:18}
 v_{e}^{(n)}(\theta_{12}) = a_{n}^{(M)} n_{M}^{2} v_{e}^{(1)} +
a_{n}^{(D)} n_{D}^{2} v_{e}^{(2)}(\theta_{12}) .
\end{equation}

Using analytic results for the angle-dependent excluded volume between
rigid $n$-mers first derived by Williamson and Jackson (WJ)\cite{21}, we
shall show that (\ref{eq:18}) is indeed satisfied exactly for
appropriate values of $a_n^{(M)}$ and $a_n^{(D)}$.  The work of WJ was
restricted to linear tangent hard-sphere chains, but it is straightforward
to generalize their analysis to LFHS chains of arbitrary bondlength $l$. 
Details are contained in the appendix.  The basic result is (see
(\ref{eq:a3})) 
\begin{equation}	\label{eq:19}
 v_{e}^{(n)}(\theta_{12}) = v_{e}^{(n)}(0) +
(n-1)^{2}v_{c}^{(2)}(\theta_{12}) ,
\end{equation}
where $v_e^{(n)}(0)$ is the excluded volume for two parallel $n$-mers
$(\theta_{12} = 0)$, given by (\ref{eq:ap1}) and (\ref{eq:a2}), and
$v_c^{(2)}(\theta_{12})$ is the contribution from the so-called ``central
region'' (in the terminology of WJ) of the overlap volume between two
diatomic molecules.  This is given by (A15) in the appendix. 
Note that, contrary to the remarks by WJ, the latter involves ``compact
and tractable'' analytic formulae. 

Before comparing (\ref{eq:18}) and (\ref{eq:19}), one rewriting of the
former is required.  As already noted, the ``reference'' monomer and dimer
fluids may be characterized by ``effective'' values of the numbers $n_M$
and $n_D$.  Similarly, it may be allowed that the diameters and
bondlengths of the reference particles differ from those of the original
chain molecule.  This feature arises in GFD theory \cite{6,26} due to the
assumption that the volume fractions of the reference fluids should equal
that of the original $n$-mer fluid, $\eta = \rho v_n$, where $v_n$ is the
$n$-mer molecular volume.  Equivalently, this requires that $v_{n} =
n_{M}v_{1} = n_{D}v_{2}$, where $v_{1}$ and $v_{2}$ are the molecular
volumes of the reference monomers and dimers, respectively\cite{33}.  The
equality of the volume fractions of the reference fluids and $n$-mer fluid
arises automatically in the case of tangent hard-sphere chains with $l =
d$, and no adjustment of the number and sizes of reference particles is
needed. This is no longer true when a chain of fused hard spheres $(l <
d)$ is decomposed into reference monomers and dimers\cite{6,26}.
Therefore, we suppose that the diameter of a reference monomer is denoted
$d_M$, the diameter of a monomer in a reference dimer is denoted $d_D$,
while the bondlength of the reference dimer is denoted $l_D$.  It is these
lengths which characterize the excluded volumes $v_e^{(1)}$ and
$v_e^{(2)}(\theta_{12})$ in (\ref{eq:18}), whereas all terms in
(\ref{eq:19}) for $v_{e}^{(n)}(\theta_{12})$ are characterized by the
diameter $d$ and bondlength $l$ of the original $n$-mer.  Hence, it is
appropriate to rewrite (\ref{eq:18}) in terms of the original molecular
dimensions.  On dimensional grounds, as confirmed by the equations in the
appendix, the dimer excluded volume scales as
\begin{equation}
\label{eq:20}
 v_{e}^{(2)}(\theta_{12}, d_{D}, l_{D}) = d_{D}^{3} \Psi\left(\theta_{12},
\frac{l_{D}}{d_{D}}\right) ,
\end{equation}
where the dependence of $v_{e}^{(2)}(\theta_{12})$ on molecular lengths
has been indicated.  Then (\ref{eq:18}) can be rewritten as
\begin{eqnarray}	\label{eq:21}
 v_{e}^{(n)}(\theta_{12}) & = & a_{n}^{(M)} n_{M}^{2}
\left(\frac{d_{M}}{d}\right)^{3} v_{e}^{(1)} + a_{n}^{(D)} n_{D}^{2}
\left(\frac{d_{D}}{d}\right)^{3} v_{e}^{(2)}(\theta_{12}) ,
\end{eqnarray}
where $v_e^{(1)}$ and $v_e^{(2)}(\theta_{12})$ now refer to the {\it
original} monomer and dimer inside the $n$-mer.  (Note $v_e^{(1)} = 4\pi
d^3/3$.) Comparing (\ref{eq:19}) and (\ref{eq:21}) now gives
\begin{equation}	\label{eq:22a}
 n_D a_n^{(D)} = \left(\frac{d}{d_D}\right)^3 \frac{(n-1)^2}{n_D} ,
\end{equation}
and
\begin{eqnarray}
 n_M a_n^{(M)} & = & \left(\frac{d}{d_M}\right)^3 \frac{[v_e^{(n)}(0)
- (n-1)^2 v_e^{(2)}(0)]}{n_M v_e^{(1)}} \nonumber\\
\label{eq:22b}
 & = & \left(\frac{d}{d_M}\right)^3 \frac{(n-2)}{n_M} 
  [2(n - 1)\left(1 - \frac{3}{4}\left(\frac{l}{d}\right)
+ \frac{1}{16}\left(\frac{l}{d}\right)^3\right)
  + (2 - 3n)] ,
\end{eqnarray}
where the second line in (\ref{eq:22b}) follows from (\ref{eq:ap1}) and
(\ref{eq:a2}) in the appendix.  Equations (\ref{eq:22a}) and
(\ref{eq:22b}) have been written in terms of the combined parameters $n_M
a_n^{(M)}$ and $n_D a_n^{(D)}$ because, as will be seen in the next
subsection, only these combinations appear in the expressions for
thermodynamic functions.  As stated at the end of the previous subsection,
$a_n^{(D)}$ and $a_n^{(M)}$ are seen to depend only on the geometric
properties of the molecules.

Of several prescriptions introduced in GFD theory \cite{6,26} to adjust
the effective parameter values, our reliance on the scaling relation
(\ref{eq:20}) restricts us to those which conserve the
reduced bondlength, $l_D/d_D = l/d \equiv l^{*}$.  These are the
approaches analogous to the A, C, and AC versions of GFD theory \cite{6}. 
However, it turns out that these methods all yield {\it identical}
thermodynamic results in the present theory.  This follows from the fact
that, according to (\ref{eq:22a}) and (\ref{eq:22b}), $n_M
a_n^{(M)}$ and $n_D a_n^{(D)}$ depend on the diameters and numbers (for
fixed $l^*$) only through the combinations $n_M (d_M / d)^3$ and $n_D (d_D
/ d)^3$, respectively, the values of which are uniquely determined by the
condition of equal volume fractions\cite{33}. 

Let us briefly compare the present expressions for $a_n^{(D)}$ and
$a_n^{(M)}$ in (\ref{eq:22a}) and (\ref{eq:22b}) with those given by GFD
theory for LFHS chains.  The latter were derived recently by Mehta and
Honnell \cite{7} using GFD-A theory, which corresponds to taking $d_M =
d_D = d$ while the effective numbers of monomers $n_M$ and dimers $n_D$
are adjusted to conserve volume fraction.  For this case, Mehta and
Honnell \cite{7} showed that the function $Y_n$ in (\ref{eq:12}) has the
value $(n - 2)$ and thus (\ref{eq:12}) reduces to
\begin{equation}	\label{eq:23}
 n_M a_n^{(M)} = -(n - 2) \quad , \quad n_D a_n^{(D)} = n - 1 \quad
(GFD-A)
\end{equation}
independent of bondlength.  For simplicity, we compare (\ref{eq:23}) with
(\ref{eq:22a}) and (\ref{eq:22b}) in the case of tangent spheres, for
which $n_M = n$ and $n_D = n/2$.  Eqs.(\ref{eq:22a}) and (\ref{eq:22b})
then give
\begin{eqnarray}	\label{eq:24}
 n_M a_n^{(M)} & = & \frac{(n - 2)(11 - 19n)}{(8n)} \longrightarrow
- \frac{19}{8}n	\nonumber\\
 n_D a_n^{(D)} & = & \frac{2(n - 1)^{2}}{n} \longrightarrow 2n ,
\end{eqnarray}
where we indicate the leading asymptotic dependence for large $n$. 
In this limit, the present parameters $n_M a_n^{(M)}$ and $n_D a_n^{(D)}$
have nearly double the magnitude of those given by GFD theory.  This
leads, via $a_n^{(D)}$, to a significantly stronger angular dependence of
the excluded volume (\ref{eq:18}) than would be predicted by GFD theory. 
One notes that $a_n^{(M)}$ is negative for $n > 2$ according to both
(\ref{eq:23}) and (\ref{eq:24}), and thus the non-ideal thermodynamic
behavior of the fluid can be considered to involve a {\it subtraction} of
the behavior of the monomer fluid from that of the dimer fluid.  It can be
shown \cite{27} that the approximate factor of two difference between GFD
and the present theory for the separate parameters $a_M a_n^{(M)}$ and
$n_D a_n^{(D)}$ largely cancels out for the second virial coefficient in
an isotropic phase, consistent with the findings of Mehta and Honnell
\cite{7}.  By construction, the present theory yields exact values of
$\hat{B}_2$ in both isotropic and nematic phases of LFHS chain fluids with
arbitrary $n$. 

\subsection{Free Energy Minimization}	\label{sec:fe}

 The total Helmholtz free energy $F$ for a uniform one-component LFHS
chain fluid follows from (\ref{eq:2}), (\ref{eq:10}), and (\ref{eq:13}): 
\begin{eqnarray}	\label{eq:25}
 F & = & kT \rho V \int d\omega f(\omega) [\ln(4\pi\Lambda\rho f(\omega))
- 1] + a_n^{(M)} \Delta F^{(M)}(\rho_M) +a_n^{(D)} \Delta 
F^{(D)}(\rho_D;[f(\omega)]) .
\end{eqnarray}
The notation in (\ref{eq:25}) for $\Delta F^{(M)}$ and $\Delta F^{(D)}$ is
meant to indicate that the former is a function of the monomer number
density $\rho_M$, while the latter is a function of $\rho_{D}$ while being
a {\it functional} of the angular distribution $f(\omega)$. 

For the isotropic phase, we shall follow previous work on GFD
theory and use the ``exact'' Carnahan-Starling \cite{28} and
Tildesley-Streett\cite{29} equations of state to obtain $\Delta F^{(M)}$
and $\Delta F^{(D)}$: 
\begin{eqnarray}	\label{eq:26}
 \frac{\Delta F^{(M)}(\rho_{M})}{VkT} & = & \rho_{M} \frac{\eta(4 -
3\eta)}{(1 - \eta)^{2}} \equiv \rho_{M} a_{CS}(\eta) ,	\\
 \frac{\Delta F^{(D)}(\rho_{D})}{VkT} & = & \rho_{D} [H'\ln(1 -
\eta) + \frac{\eta}{2(1 - \eta)^{2}}[2(F' + H') - (F' - G' + 3H')\eta]]
\nonumber\\ \label{eq:27a}
 & \equiv & \rho_{D} a_{TS}(\eta) ,
\end{eqnarray}
where
\begin{eqnarray}
 F' & = & F + 3 = 4 + 0.37836l^{*} + 1.07860(l^{*})^{3} ,
\nonumber\\
 G' & = & G - 3 = -2 + 1.30376l^{*} + 1.80010(l^{*})^{3} ,
\nonumber\\ \label{eq:27b}
 H' & = & H - 1 = 2.39803l^{*} + 0.35700(l^{*})^{3} ,
\end{eqnarray}
(and where F, G, H denote the same quantities as defined by Tildesley and
Streett \cite{29}).  As stated earlier, the volume fraction $\eta$ is
assumed to be the same for the $n$-mer, dimer, and monomer fluids.  In a
uniform nematic fluid, the excess monomer free energy is still given by
(\ref{eq:26}).  The simplest conceivable ansatz for $\Delta F^{(D)}
(\rho_D;[f(\omega)])$ is that suggested by decoupling theory 
\cite{18,19,30}, namely
\begin{equation}	\label{eq:28a}
 \Delta F^{(D)}(\rho_D;[f(\omega)]) = \Delta F^{(D)}(\rho_D) J[f(\omega)]
,
\end{equation}
where J is given by
\begin{equation}	\label{eq:28b}
 J[f(\omega)] = \frac{\int d\omega_1 d\omega_2 f(\omega_1)
f(\omega_2) v_e^{(2)}(\theta_{12})}{\int
d\omega_1 d\omega_2 v_e^{(2)}(\theta_{12})/(4\pi)^2} .
\end{equation}
From (\ref{eq:16}), the functional $J[f(\omega)]$ is equivalent to the
ratio between the second virial coefficients of the nematic and isotropic
dimer fluids. 

A self-consistent equation determining the equilibrium form of the angular
distribution function follows by functional minimization of $F$ with
respect to $f(\omega)$, subject to the normalization condition $\int
d\omega f(\omega) = 1$.  We obtain
\begin{equation}	\label{eq:29a}
 f(\omega) = \frac{e(\omega)}{\int d\omega' e(\omega')} ,
\end{equation}
where
\begin{equation}	\label{eq:29b}
 e(\omega_1) = \exp \left[-2n_Da_n^{(D)}a_{TS}(\eta) \frac{\int
d\omega_2 f(\omega_2) v_e^{(2)}(\theta_{12})}{ 
\int d\omega_1 d\omega_2 v_e^{(2)}(\theta_{12})/(4\pi)^2}\right]
\end{equation}
and we have used the fact that $\rho_D = n_D\rho$.  The excluded volume
$v_{e}^{(2)}(\theta_{12})$ in (\ref{eq:28b}) and (\ref{eq:29b}) is that
for reference dimers of diameter $d_D$ and bondlength $l_D$, but explicit
factors of $d_D^3$ cancel in the ratio of integrals due to the scaling law
(\ref{eq:20}) and only the reduced bondlength $l^{*}$ is required in
specifying those integrals. 

For a given volume fraction $\eta$, the integral equation (\ref{eq:29a}) 
is solved numerically by iteration as in ref.\cite{30}.  To evaluate the
angular integrals involving $v_e^{(2)}(\theta_{12})$ in (\ref{eq:29b}), we
express the molecular orientation $\omega_{i} \equiv (\theta_i,\phi_i)$ in
the director frame of reference, $i.e.$, with polar angle $\theta_i$
measured from the nematic director axis.  Then $f(\omega_i) = f(\theta_i)$
is independent of the azimuthal angle $\phi_i$.  The angle $\theta_{12}$
between the two rod axes can be expressed as
\begin{equation}	\label{eq:30}
 \cos \theta_{12} = \sin \theta_1 \sin \theta_2 \cos \phi_2 + \cos
\theta_1 \cos \theta_2 ,
\end{equation}
(where we arbitrarily set $\phi_1 = 0$).  Then
\begin{equation}	\label{eq:31a}
 \int d\omega_2 f(\omega_2) v_e^{(2)}(\theta_{12}) = \int_{0}^{\pi}
\sin \theta_2 d\theta_2 f(\theta_2)
\overline{v}_e^{(2)}(\theta_1,\theta_2) ,
\end{equation}
where
\begin{equation}	\label{eq:31b}
 \overline{v}_e^{(2)}(\theta_1,\theta_2) = \int_{0}^{2\pi} d\phi_2
v_e^{(2)}(\theta_{12})~ .
\end{equation}

Using the analytic formulae for $v_e^{(2)}(\theta_{12})$ from the
appendix, the integrations over $\theta_2$ and $\phi_2$ in (\ref{eq:31a}) 
and (\ref{eq:31b}) are performed numerically by the trapezoid rule (for
each pair of $\theta_1$ and $\theta_2$, the integration over $\phi_2$ has
to be done only once).  The variables $\cos \theta$ and $\phi$ are
discretized on grids of typical stepsizes $\Delta (\cos \theta) = 0.005$
and $\Delta \phi = \pi/200$.  Changing the frame of reference to relative
orientations $(\theta_{12},\phi_{12})$ could be applied to reduce the
number of integration variables involved in the denominator of the
exponential in (\ref{eq:29b}), but for numerical consistency and with a
view towards cancellation of errors, that integral is evaluated here in a
manner analogous to (\ref{eq:31a}) (noting that the integration $\int
d\omega_1/(4\pi)$ becomes redundant).  Similar numerical integration
techniques are used to evaluate $J[f(\omega)]$, the denominator of
(\ref{eq:29a}) and the order parameter $S_2$ defined by (\ref{eq:60}) in
Section \ref{sec:lths}. 

Once the numerical solution of (\ref{eq:29a}) is obtained, the corresponding
equilibrium thermodynamic potentials can be evaluated from the Helmholtz
free energy given by eqs.(\ref{eq:25})-(\ref{eq:28b}).  Substituting
(\ref{eq:29a}) into the logarithm of (\ref{eq:25}), one obtains
\begin{equation}	\label{eq:32}
 \frac{F}{kTV} = \rho \left[\ln\left(\frac{4\pi\rho\Lambda}{\int
d\omega e(\omega)}\right) - 1\right] + a_n^{(M)}\rho_M a_{CS}(\eta) 
 - a_n^{(D)}\rho_D a_{TS}(\eta)J[f(\omega)] .
\end{equation}
The chemical potential $\mu$ follows from
\begin{equation}
 \mu = \left(\frac{\partial(F/V)}{\partial\rho}\right)_T
.	\nonumber
\end{equation}
Hence
\begin{eqnarray}	\label{eq:33}
 \frac{\mu}{kT} & = & \ln\left[\frac{4\pi\rho\Lambda}{\int d\omega
e(\omega)}\right] + n_M a_n^{(M)}\frac{\Delta\mu_M(\eta)}{kT}
 + n_D a_n^{(D)}J[f(\omega)]\left[\frac{\Delta\mu_D(\eta)}{kT} -
2a_{TS}(\eta)\right] ,
\end{eqnarray}
where $\Delta\mu_M$ and $\Delta\mu_D$ are the excess chemical
potentials of the monomer and (isotropic) dimer fluids, respectively:
\begin{eqnarray}	\nonumber
 \frac{\Delta\mu_M(\eta)}{kT} & = & a_{CS}(\eta) + \eta a'_{CS}(\eta) ,
\\
 \frac{\Delta\mu_D(\eta)}{kT} & = & a_{TS}(\eta) + \eta a'_{TS}(\eta) ,
\label{eq:34}
\end{eqnarray}
where $a'(\eta) \equiv da(\eta)/d\eta $.  Finally, the equilibrium
pressure is obtained from $P = \rho\mu - F/V$, which gives
\begin{equation}	\label{eq:35}
 \frac{P}{kT} = \rho + \rho_M a_n^{(M)}\Delta Z_M(\eta) +
\rho_D a_n^{(D)}J[f(\omega)]\Delta Z_D(\eta) ,
\end{equation}
where $\Delta Z_M$ and $\Delta Z_D$ are the excess monomer and (isotropic)
dimer compressibility factors, respectively, given by
\begin{eqnarray}
 \Delta Z_M(\eta) & = & \eta a'_{CS}(\eta) = \frac{2\eta(2 - \eta)}{(1 -
\eta)^{3}} ,	\nonumber\\
\Delta Z_{D}(\eta) & = & \eta a'_{TS}(\eta) = \frac{\eta(F' + G'\eta -
H'\eta^{2})}{(1 - \eta)^{3}} .	\label{eq:36}
\end{eqnarray}
The isotropic limits of the preceding equations are obtained by setting
$f(\omega) = 1/(4\pi)$.  In this limit, $J[f(\omega)] = 1$.
Coexistence between isotropic(I) and nematic(N) phases is then evaluated
by solving the equations $P(\eta_I) = P(\eta_N)$ and $\mu(\eta_I) =
\mu(\eta_N)$, which is done by a Newton-Raphson procedure.

\section{Results}	\label{sec:results}
\subsection{LTHS Chains}	\label{sec:lths}

 We consider first the case of tangent hard-sphere $n$-mers, before
examining $n$-mers with arbitrary bondlength.  The primary limitation to
our analysis of the theory is the lack of extensive simulation data
available for comparison.  Of the Monte Carlo studies which have examined
hard-sphere chains, the majority have focused on semi-flexible systems of
tangent chains.  Several of these studies have dealt with the limit of
infinite rigidity and therefore yield results which can be used here for
comparison.  In particular, the recent studies by Yethiraj and
Fynewever\cite{17b,17a} provide us with substantial analyses of the
$8$-mer and $20$-mer LTHS chains.  Williamson and Jackson\cite{16}
recently examined $7$-mer LTHS chains and performed comparisons with
theoretical treatments involving the exact excluded volume for the
chains\cite{21}. These studies will be the primary sources for comparisons
with our results. 

The conventional orientational order parameter $S_{2}$ is defined as
\begin{equation}	\label{eq:60}
 S_{2} = \int P_{2}(\cos\theta)f(\omega)d\omega ,
\end{equation}
where $P_{2}$ is the second Legendre polynomial.  This parameter has the
limits of $S_{2} = 0$ for an isotropic phase and $S_{2} = 1$ for a
perfectly aligned nematic phase. Figure~1 (a) and (b) show the order
parameter $S_{2}$ as a function of volume fraction $\eta$ for $8$-mers and
$20$-mers, respectively.  The Monte Carlo simulation data\cite{17b} is
obtained using both constant-pressure (NPT) and constant-volume (NVT)
methods, which are seen to be in very good agreement.  This data is
compared with the present theory and with the Parsons theory\cite{18}, as
employed by Yethiraj and Fynewever\cite{17b,17a} (using a simple {\it
hard-sphere} reference fluid and incorporating the excluded volume derived
by Williamson and Jackson\cite{21}).

In general, the present theory is in very good agreement with the
simulation data for the degree of ordering in each system.  The order
parameter at which the nematic phase first appears is of the order $S_{2}
> 0.5$. The degree of ordering continues to increase dramatically over an
extremely short density range, until $S_{2} \approx 0.8$, after which the
value of the order parameter begins to level off.  This general behavior
is also evident in the Parsons theory, although the values obtained for
$S_{2}$ and the coexistence densities are less accurate.  For the $8$-mer
LTHS chain system, the values of the order parameter predicted by our
theory are in excellent agreement with the values obtained by simulation. 
Our theoretical results for the $20$-mers exhibit slightly lower values of
$S_{2}$ in comparison with the simulations, although the amount of data
available is much smaller in this case.  The Parsons theory systematically
underestimates the degree of ordering in the nematic phase, as is
evidenced in Figs.~1(a) and (b). 

In Figs.~2(a) and (b), the reduced pressure $P^{*} = Pv_{n}/kT$ is plotted
as a function of volume fraction for the $8$-mer and $20$-mer LTHS chains,
respectively.  The simulation data clearly show the coexistence densities
between the isotropic and nematic branches.  The values for these
densities as yielded by both theory and simulation are given in Table~1,
along with the order parameter $S_{2}$ and $P^*$ in the coexisting nematic
phase.  It is seen that the coexistence regions predicted by the present
theory agree very well with simulation, although the reduced pressure at
any density shows poor agreement.  The present theory overestimates the
value of $P^{*}$ throughout the density range, a discrepancy which becomes
more pronounced at higher densities.  On the other hand, the Parsons
theory predicts coexistence densities substantially higher than those
observed in simulation, as well as a larger coexistence range, as is
depicted on both graphs.  The Parsons theory underestimates the values of
the reduced pressure as a function of density for both systems.  Yethiraj
and Fynewever\cite{17b} note that SPT treatments also underestimate the
pressure for the $8$-mer system, while overestimating it for larger
molecules.

The $7$-mer LTHS chain has been recently examined by Williamson and
Jackson [WJ] \cite{16} in light of the explicit calculation of the
excluded volume for LTHS chains\cite{21}.  WJ performed Monte Carlo
simulations for a system of $N = 576$ molecules.  The results of the
simulation were compared with several theories which were modified to
incorporate the calculation of the exact excluded volume.  In particular,
WJ compare the simulation results with those obtained from a modified
version of the Vega and Lago theory\cite{15}.  The original Vega and Lago
theory is based on the form of the decoupling approximation in
(\ref{eq:6}) and (\ref{eq:7}), where the reference fluid is the isotropic
phase of the $n$-mer fluid.  WJ modify the original Vega and Lago theory
to incorporate their exact analytic calculation of the excluded volume
into $\hat{B}_{2}$, and obtain the isotropic phase free energy from
TPT\cite{1,34}.  In Fig.~3, we compare the results for $S_{2}$ from the
present theory with the Monte Carlo simulation data.  As in Fig.~1, the
present theory predicts the order parameter with excellent accuracy.  The
general behavior of $S_{2}$ is similar to the previous graphs.  The
comparison of reduced pressure versus volume fraction for the $7$-mer
system is made in Fig.~4. The modified Vega and Lago theory yields
accurate pressures in comparison with simulation, but the predicted
transition densities (evidenced by the plateau in the trace)  are too low. 
This is also indicated in Table~1, which shows that the present theory
yields coexistence densities lying within the ranges obtained through the
simulations.  The Vega-Lago theory also overestimates the value of $S_{2}$
in the nematic phase.  However, as in Fig.~2, the values of reduced
pressure given by the present theory clearly exceed the reported
simulation data.  This discrepancy appears to be systematic in the theory
and will be discussed in Section \ref{sec:conclude}.  It should be noted
here that the WJ simulation data predict a smectic phase to occur at
volume fractions greater than $0.37$.

The relationship between the length of molecules and the range of the
coexistence region is shown in Fig.~5.  The general trend is an
increasing difference between the densities of the isotropic and nematic
branches at coexistence as the number of monomers constituting the
molecule increases.  This trend is consistent with simulation, as is shown
in this figure for $7$-mers, $8$-mers and $20$-mers.  The Parsons theory
and various SPT treatments have yielded a similar trend\cite{17b},
although these theories tend to overestimate the width of the coexistence
region. In addition, the values of the coexistence densities obtained from
these theories are in poor agreement with simulation, while the present
theory provides a quantitatively accurate description of these densities.

\subsection{LFHS Chains}	\label{sec:lfhs}

 Simulations of LFHS chains have been performed by Whittle and
Masters\cite{32} for the cases of $6$-mers with a reduced bondlength of
$l^{*} = 0.5$ and $8$-mers with $l^{*} = 0.5$ and $0.6$.  These three
systems were recently investigated by Mehta and Honnell\cite{7} using GFD
theory, comparing their results with the Whittle and Masters\cite{32}
simulation data and with a modification of TPT\cite{1}. The GFD theory and
the modified TPT are unable to treat ordering, and therefore are
inappropriate at densities above the isotropic-nematic transition which
was evidenced in the simulation of the $8$-mer system with $l^{*} = 0.6$. 
In Fig.~6, we perform an analogous study of the three fluids, comparing
the previous results\cite{7} with those of the present theory.
Figure~6(a) shows the reduced pressure as a function of volume fraction
for the $6$-mer LFHS chains with $l^{*} = 0.5$.  There is no evidence of a
nematic transition in the simulations, nor is there any indication through
the theoretical treatments.  The present theory and the modified TPT both
overestimate the values of the reduced pressure, while the GFD theory
appears to predict the reduced pressure to a fair degree of accuracy. This
trend is also apparent in Figs.~6(b)  and (c)  for the $8$-mer cases.  In
the latter plots, the present theory is seen to predict a stable nematic
branch at sufficiently high volume fractions.  The simulation data in
Fig.~6(b) exhibit no such transition, and we are unaware of any other
simulations which have been done for this system.  It should be noted that
the monomeric reference fluid does not exist as a stable fluid above $\eta
\approx 0.494$\cite{7} and as such the theory is probably invalid at these
high densities.

Figure~6(c) considers the $8$-mer fluid with $l^{*} = 0.6$, a system which
clearly exhibits nematic ordering.  At first glance it appears that the
present theory, while succeeding in predicting a stable nematic branch,
predicts that it occurs at volume fractions much greater than the
simulations indicate.  However, it should be noted that although Whittle
and Masters\cite{32} report a nematic branching at $\eta \approx 0.33$,
the order parameter was not found to be stable until much higher volume
fractions, $\eta \ge 0.419$.  The predictions of the present theory for
transition properties are compared with the simulation results in Table~1.
Once again, we see that there is good agreement for the value of the
nematic order parameter $S_2$.

It is of interest to examine the present theory in the ``spherocylinder''
limit, which is obtained in the limits $n \rightarrow \infty$, $l
\rightarrow 0$, such that $L \equiv (n-1)l$ remains finite. As discussed
in the appendix, our analytical result for the LFHS excluded volume
reduces (as should be expected) to that for hard spherocylinders of
cylinder length $L$ in this limit.  Figure~7 shows the nematic and
isotropic coexistence volume fractions for LFHS $n$-mers as a function of
$n$, for a fixed value of $L/d = 19$, in comparison with the values for
spherocylinders obtained from the Lee theory\cite{19} and from Monte Carlo
calculations (using the Gibbs-Duhem integration procedure) by Bolhuis and
Frenkel\cite{bf}.  One sees that both the individual volume fractions and
their differences increase slowly with $n$, and approach the asymptotic
spherocylinder values reasonably well, although underestimating the
coexistence width in the limit.  In more detail, it can be shown that the
present theory in the spherocylinder limit yields exactly the same angular
distribution function $f(\omega)$, at a given volume fraction $\eta$, as
in the Lee theory\cite{19}.  However, the free energy and pressure are
actually predicted to {\it diverge} in this limit, which accounts for the
narrower coexistence width found in Fig.~7. 

\section{Concluding Remarks}	\label{sec:conclude}

 In this paper, we have introduced two key modifications into the
generalized Flory-dimer (GFD) theory to describe nematic behavior in
hard-sphere chain fluids of arbitrary intramolecular bondlength.  The
first modification is the inclusion of the exact excluded volume and
second virial coefficient for LFHS chain molecules, based on generalizing
the earlier calculation of Williamson and Jackson\cite{21}.  This
procedure results in mixing parameters for the reference monomer and dimer
fluids which are dependent only on the geometric properties of the
molecules (see (\ref{eq:22a}) and (\ref{eq:22b})).  A related feature is
that the ansatz of equal volume fractions {\it uniquely} determines the
values of the relevant combinations of the effective reference parameters
$n_D, d_D, n_M$ and $d_M$, in contrast to previous studies based on GFD
and related theories\cite{4,6,26}.  The second modification of GFD theory,
in order to account for the possibility of nematic ordering in the system,
is the weighting of the excess dimer free energy by the ratio between the
second virial coefficients of the nematic and isotropic dimer fluids (see
(\ref{eq:28a}) and (\ref{eq:28b})).  This ansatz, suggested by decoupling
theory, is dependent on the angular distribution function, which is
determined self-consistently. 

The present theory is found to be in excellent agreement with simulations
of LTHS fluids in determining coexistence densities and the nematic order
parameter as a function of density.  As is seen clearly in Table~1 and
Fig.~5, the transition densities for the $7$-mer, $8$-mer and $20$-mer
fluids predicted by the theory fall within the coexistence range found in
simulations.  In particular, the $7$-mer system studied by Williamson and
Jackson\cite{16} shows much better agreement with the present theory than
with the Vega and Lago\cite{15} and the Parsons\cite{18} theories for the
coexistence densities as well as the nematic order parameter at the
transition. The present theory is able to account for stable nematic
branches in LFHS chains of sufficient length, for which theoretical work
has been lacking up to now.  The $8$-mer with $l^{*} = 0.6$ simulated by
Whittle and Masters\cite{32} yielded transitions at lower densities than
predicted by the present theory, although the simulated nematic branch did
not become stable until densities similar to those predicted in our work.  
In addition, transitions between the isotropic and nematic phases are
predicted by the present theory for several systems which have yet to be
simulated (see Figs.~5 and 7).  The agreement between the present theory
and simulations is promising and it is hoped that further work will be
encouraged.

For all systems studied, the present theory yields values for the pressure
which exceed those given by available simulation data as well as previous
theories.  This discrepancy is not accounted for within this paper, and
further work needs to be done to resolve the problem.  It was noted in
Section \ref{sec:lfhs} that the free energy and pressure given by the
present theory actually diverge in the spherocylinder limit $n \rightarrow
\infty$, $l \rightarrow 0$, with $(n-1)l \equiv L$ finite.  This
divergence can be traced to the incorrect limiting behavior of the
Tildesley-Streett (TS)\cite{29} dimer equation of state for small reduced
bondlength $l^{*}$. The TS equation of state predicts that the reduced
isotropic second virial coefficient (as well as higher virial
coefficients) varies linearly with $l^{*}$ (see (\ref{eq:27b})), in
contrast with the correct leading-order variation proportional to
$(l^{*})^2$ (see (\ref{eq:a17})).  However, use of other dimer equations
of state which do have the correct dependence for $l^* \rightarrow 0$,
such as the ``improved scale particle theory'' of Boublik and
Nezbeda\cite{bn}, produces negligible differences from most of the results
described in this paper, apart from those shown in Fig.~7 for $n \ge 40$.

Despite the mixed agreement of our results with those of available
simulations, we believe that the approach described here has advantages
over other current density functional theories of chain
fluids\cite{2a,2b,17b,13,14a,14b,15,16,17a}, particularly for considering
extensions of the theory to non-uniform fluids and ones containing
semi-flexible molecules.  In Section \ref{sec:gen}, we briefly indicated
these generalizations of the theory, which we are currently investigating
in more detail.  Our approach retains the geometrically motivated spirit
of the GFD and TPT theories in utilizing reference fluids composed of
monomer and dimer subunits, whose properties can be readily determined.
This contrasts with the form of the decoupling approximation used in
refs.\cite{15,16}, requiring {\it $\grave{a}$ priori} information about
the thermodynamics of the isotropic phase of the full system being
considered, which may be either unavailable or computationally difficult
to obtain in the more general cases.  At the same time, the dimer 
reference fluid has the crucial property of involving orientational
degrees of freedom.  This feature corrects a limitation of previous
density-functional approaches for non-uniform chain
fluids\cite{2a,2b,13,14a,14b} based on a purely monomeric reference fluid,
which are unable to account for orientational ordering effects.

\textbf{Acknowledgments.} This work was supported by a research grant from
the Natural Sciences and Engineering Research Council (Canada), and by a
NATO Collaborative Research Grant.  We are grateful to Drs. B.G.~Nickel
and C.G.~Gray for their insightful comments.

\appendix
\setcounter{section}{0}
\renewcommand{\thesection}{}
\section{Excluded volume of LFHS chains}
\setcounter{equation}{0}
\renewcommand{\theequation}{A\arabic{equation}}
 Williamson and Jackson (WJ)\cite{21} recently derived an expression for
the excluded volume between a pair of LTHS $n$-mers at an arbitrary
relative orientation $\theta_{12}$. Here we extend their analysis to LFHS
$n$-mers of arbitrary bondlength.

The excluded volume for parallel orientation, $\theta_{12} = 0$, is a
straightforward extension of the WJ result, and is described by a
chain of $(2n - 1)$ overlapping spheres of radius $d$ and volume
$v_{S} = 4\pi d^3/3$ separated by the $n$-mer bondlength $l$.  The volume
is given by the relation
\begin{equation}	\label{eq:ap1}
 v_{e}^{(n)}(0) = (2n - 1)v_{S} - 2(n - 1)v_{o} .
\end{equation}
The overlap volume between a pair of adjacent spheres, $v_{o}$, is dependent on
the bondlength-to-diameter ratio $l/d$, and is given by
\begin{equation}	\label{eq:a2}
 v_{o} = \frac{4\pi}{3} d^{3}\left[1 - \frac{3}{4}\left(\frac{l}{d}\right)
+ \frac{1}{16}\left(\frac{l}{d}\right)^{3}\right] .
\end{equation}

For a pair of $n$-mer molecules in an arbitrary orientation 
$\theta_{12}$, the excluded volume is represented diagrammatically as
$n^2$ overlapping spheres of radius $d$ whose centers lie on the corners
of a rhombus (Fig.~8).
The centers of adjacent spheres are separated by $l$ and the rhombus has
angle $\theta_{12}$.
We define the ``central region'' (in the terminology of WJ) as the
parallelpiped based on the rhombus, taken to lie in the $xy$ plane, and
extending along the $z$ axis to distances $\pm d$. As in WJ\cite{21}, it can
be shown that the
excluded volume outside the central region is equal to the excluded volume
in the
$\theta_{12} = 0$ orientation.  Furthermore, one can
show that the excluded volume contained within the central region for an
$n$-mer is
$(n - 1)^{2}$ times that of a dimer.  The total excluded volume is thus
given by
\begin{equation}	\label{eq:a3}
 v_{e}^{(n)}(\theta_{12}) = v_{e}^{(n)}(0) +
(n-1)^{2}v_{c}^{(2)}(\theta_{12}) ,
\end{equation}
where $v_e^{(n)}(0)$ is given by (\ref{eq:ap1}) and 
$v_c^{(2)}(\theta_{12})$ is the contribution from the central region to
the excluded volume between two diatomic molecules.

Following WJ \cite{21}, it is convenient to evaluate $v_c^{(2)}(\theta_{12})$
by considering infinitely thin slices parallel to $xy$ at varying heights
$z$. Due to mirror symmetry in the $xy$ plane, we need to examine only
distances $0 \le z \le d$. There are three distinct ranges of $z$ which
must be considered.  The outermost range of $z$ is characterized by the
absence of two-body overlaps between the circular cross-sections through 
the spheres. The total excluded area of a slice
parallel to $xy$ within this range is that of a circle,
\begin{equation}	\label{eq:a4}
 A_c^I(z) = \pi(d^2 - z^2),
\end{equation}
where z ranges from $d$ to $h_{II}$, the height at which two-body
overlaps begin to occur.  The distance $h_{II}$ is dependent on the
magnitude of the angle between the molecules, $\theta_{12}$.  For large
enough $\theta_{12}$ ($\pi/3 \le \theta_{12} \le \pi/2$), overlap between
the circular cross-sections first occurs for adjacent pairs of spheres.
At this height, the radius of a circular cross-section is $r = l/2$, and
the corresponding value of $z = h_{II}$ is:
\begin{equation}	\label{eq:a6}
 h_{II}^{(1)} = \sqrt{d^2 - \left(l/2\right)^2} .
\end{equation}
For smaller $\theta_{12}$ ($0 \le \theta_{12} \le \pi/3$), the first
two-body overlaps occur between opposing spheres (Fig.~9).  This height
is given by the relation
\begin{equation}	\label{eq:a7}
 h_{II}^{(2)} = \sqrt{d^2 - l^2 \sin^2\left(\theta_{12}/2\right)} .
\end{equation}
For both angular cases evaluated above, the region of two-body overlaps
extends until $z = h_{III}$, at which height the circular 
cross-sections through the spheres begin overlapping in a three-body
configuration.  The value for this height is given by
\begin{equation}	\label{eq:a8}
 h_{III} = \sqrt{d^2 - \left(\frac{l}{2 
\cos\left(\theta_{12}/2\right)}\right)^2} .
\end{equation}
Throughout the range $h_{III} < z < d$, the total excluded area of a
parallel slice can be expressed as
\begin{equation}	\label{eq:a9}
 A_c^{II}(z) = A_c^I(z) - A_o(z)
\end{equation}
where $A_c^I(z)$ is given by (\ref{eq:a4}) while $A_{o}(z)$ is the total
two-body overlap area between the circles at height $z$.  This overlap
area has distinct contributions depending on whether the overlapping
spheres are adjacent or opposite from each other (Fig.~9), and is given
by
\begin{equation}	\label{eq:a10}
 A_o(z) = 2a_o(l,z) + a_o(\hat{l},z) ,
\end{equation}
where $\hat{l} = 2l \sin\left(\theta_{12}/2\right)$.  Here,
$a_o(l',z)$ is the area of overlap between a pair of circles of radius
$r$, separated by $l'$, given by
\begin{equation}	\label{eq:a11}
 a_o(l',z) = 2\left[r^2\cos^{-1}\left(\frac{l'}{2r}\right) -
\left(\frac{l'}{2}\right)\left(r^2 - 
\frac{l'^2}{4}\right)^{\frac{1}{2}}\right] ,
\end{equation}
keeping in mind that $r = \sqrt{d^2 - z^2}$.  The relation
(\ref{eq:a11}) differs from equation (7) of WJ by a factor of 2 because WJ
calculate only half of the overlap area.  Remembering that the
contributions of the two types of overlap begin at different values of $z$,
the total contribution to the central excluded volume from regions $I$ and
$II$ is
\begin{eqnarray}	\nonumber
 v_c^{I,II}(\theta_{12}) & = & 2 \int_{h_{III}}^{d} dz\ A_c^{II}(z) \\
& = & 2\pi \left[d^2\left(d - h_{III}\right) - \frac{1}{3}\left(d^3 -
h_{III}^3\right)\right] \nonumber \\
&& - 4\int_{h_{III}}^{h_{II}^{(1)}} dz\ a_o(l,z)
 - 2\int_{h_{III}}^{h_{II}^{(2)}} dz\ a_o(\hat{l},z) \label{eq:a12}
\end{eqnarray}
where the additional factor of 2 in (\ref{eq:a12}) is due to mirror
symmetry.

The integrals involving $a_{o}'s$ are quoted by WJ as being
``intractable'', but we have derived a compact and tractable form for
the indefinite integral:
\begin{eqnarray}
 \int dz'\ a_o(l',z') & = & 2\left(d^2z -
z^3/3\right)\cos^{-1}\left(\frac{l'/2}{\sqrt{d^2 - 
z^2}}\right)
 - \frac{2}{3}l'z\left[d^2 - z^2 -
\left(l'/2\right)^2\right]^{1/2}	\nonumber\\
&& - l'\left[d^2 - \frac{1}{3}\left(l'/2\right)^2\right]
\sin^{-1}\left(\frac{z}{\sqrt{d^2 - \left(l'/2\right)^2}}\right) 
\nonumber\\	\label{eq:a13}
&& + \frac{4}{3}d^3 \tan^{-1}\left[\frac{l'z/(2d)}{\left[d^2 - z^2 -
\left(l'/2\right)^{2}\right]^{1/2}}\right] .
\end{eqnarray}
One can verify that the z-derivative of the right-hand side of
(\ref{eq:a13}) equals $a_o(l',z)$.

The innermost part of the central region is defined by $z \le h_{III}$,
where three-body overlaps occur.  The whole volume of this region is
excluded, and thus its contribution from this region to $v_c^{(2)}$ is
equal to
\begin{equation}	\label{eq:a14}
 v_c^{III}(\theta_{12}) = 2l^2d\left[\sin^2\theta_{12} -
\left(\frac{l}{d}\right)^2\sin^2\left(\theta_{12}/2\right)
\right]^{1/2} .
\end{equation}
The total excluded volume of the central region is then
\begin{equation}	\label{eq:a15}
 v_c^{(2)}(\theta_{12}) = v_c^{I,II}(\theta_{12}) + v_c^{III}(\theta_{12})
,
\end{equation}
where the various contributions are given by (\ref{eq:a12}) to 
(\ref{eq:a14}) for $\theta_{12} \le \pi/2$.  Due to chain inversion
symmetry, the excluded volume for $\theta_{12} > \pi/2$ is given by
$v_c^{(2)}(\theta_{12}) = v_c^{(2)}(\pi - \theta_{12})$.  Substituting the
expressions for $h_{II}^{(1)}, h_{II}^{(2)}$ and $h_{III}$ from 
(\ref{eq:a6}), (\ref{eq:a7}) and (\ref{eq:a8}) into (\ref{eq:a12}) and
(\ref{eq:a13}), and applying several trigonometric identities (most
importantly the relation $\tan^{-1}(A) - \tan^{-1}(B) = \tan^{-1}[(A -
B)/(1 + AB)]$), we are able to reduce the result for
$v_c^{(2)}(\theta_{12})$ to the following compact form:
\renewcommand{\theequation}{A15a}
\begin{eqnarray}	\nonumber
v_c^{(2)}(\theta_{12}) & = & \left(\frac{2}{3}l^2d\right)
\sin\left(\theta_{12}/2\right) S(\theta_{12}) \\ \nonumber
&& + 4l\left[d^2 - \frac{1}{3}\left(\frac{l}{2}\right)^2\right]
\tan^{-1}\left[\frac{l}{d}
\frac{\sin\left(\theta_{12}/2\right)}{S(\theta_{12})}\right] \\ \nonumber
&& + 4l\sin\left(\theta_{12}/2\right)\left[d^2 -
\frac{l^2}{3}\sin^2\left(\theta_{12}/2\right)\right]
\tan^{-1}\left[\frac{l}{d} \frac{\cos(\theta_{12})}{S(\theta_{12})}\right]
\\ \label{eq:abjn}
&& - \left(\frac{8}{3}d^3\right) \tan^{-1}\left[
\frac{l^2\sin\left(\theta_{12}/2\right)S(\theta_{12})}{4d^2 -
l^2\left(1 + 2\sin^2\left(\theta_{12}/2\right)\right)}\right] ,
\end{eqnarray}
where
\renewcommand{\theequation}{A15b}
\begin{eqnarray}	\label{eq:abjn2}
S(\theta_{12}) & = & \left[4\cos^2\left(\theta_{12}/2\right) -
\frac{l^2}{d^2}\right]^{1/2} .
\end{eqnarray}

\setcounter{equation}{15}
\renewcommand{\theequation}{A\arabic{equation}}
It can be verified that the preceding result for the angle-dependent
excluded volume of LFHS $n$-mers reduces to that for hard spherocylinders
in the limit that $n \rightarrow \infty$, $l \rightarrow 0$, and $L \equiv
(n-1)l$ remains finite, where $L$ is identified with the cylinder length.
In this limit, (\ref{eq:ap1}) and (\ref{eq:a2}) give
\begin{equation}	\label{eq:a16}
 v_e^{(n)}(0) \rightarrow \frac{4 \pi d^3}{3}\left(1 + \frac{3}{2}
\frac{L}{d} \right) \equiv 8v_{SC} ,
\end{equation}
where $v_{SC}$ is the spherocylinder molecular volume.  From
(\ref{eq:abjn},b), one finds readily that
\begin{equation}	\label{eq:a17}
 v_c^{(2)}(\theta_{12}) = 2l^2d\sin\theta_{12} + O(l^4) .
\end{equation}
The total excluded volume (\ref{eq:a3}) then becomes
\begin{equation}	\label{eq:a18}
 v_e^{(n)}(\theta_{12}) \rightarrow 8v_{SC} + 2L^2d\sin\theta_{12} ,
\end{equation}
which agrees with the result derived for hard spherocylinders\cite{19,20}.

\pagebreak
\setcounter{table}{0}
\renewcommand{\thetable}{\arabic{table}}
\begin{table}[!h]
\caption{Coexistence results from simulation and theory.}
\begin{center}
\begin{tabular}[!h]{l|l|cccc} \hline
LFHS Chain & Source & $\eta(iso)$ & $\eta(nem)$ & $S_{2}(nem)$ & $P^*$ \\
\hline
$7$-mer, $l^{*}=1.0$ & Present theory & 0.2903 & 0.2989 & 0.6491 & 4.94\\
& MC-NPT data\cite{16} & 0.266-0.303 & 0.285-0.312 & 0.64-0.66 &
3.15-3.78\\
& Vega-Lago theory\cite{16} & 0.255 & 0.273 & $\approx$ 0.7 & 2.78\\
& Parsons theory\cite{16} & 0.303 & 0.319 & N/A & 3.12\\
\hline
$8$-mer, $l^{*}=1.0$ & Present theory & 0.2601 & 0.2689 & 0.6538 & 3.95\\
& MC-NPT data\cite{17b} & 0.257 & 0.271 & $\approx$ 0.7 & 2.65\\
& Parsons theory\cite{17b,17a} & 0.280 & 0.305 & $\approx$ 0.7 & 2.6\\
\hline
$20$-mer, $l^{*}=1.0$ & Present theory & 0.1158 & 0.1243 & 0.6947 & 0.97\\
& MC-NPT data\cite{17b} & 0.105 & 0.120 & $\approx$ 0.7 & 0.62\\
& Parsons theory\cite{17b,17a} & 0.115 & 0.140 & $\approx$ 0.75 & 0.69\\
\hline
$8$-mer, $l^{*}=0.5$ & Present theory & 0.4687 & 0.4768 & 0.6168 & 13.1\\
& MC-NPT data\cite{32} & \multicolumn{3}{c}{No transition found} & \\
\hline
$8$-mer, $l^{*}=0.6$ & Present theory & 0.4167 & 0.4251 & 0.6224 & 9.8\\
& MC-NPT data\cite{32} & & 0.419 & 0.624 & 5.7\\
\hline
\end{tabular}
\end{center}
\end{table}

\begin{figure}[!h]
\caption{Variation of order parameter $S_{2}$ with volume fraction $\eta$,
comparing present theory with the Parsons theory and Monte Carlo
data[11,21], for (a)$8$-mer LTHS chains, and (b)$20$-mer LTHS
chains.}
\end{figure}
\begin{figure}[!h]
\caption{Variation of the reduced pressure with volume fraction $\eta$,
comparing present theory with the Parsons theory and Monte Carlo
data[11,21], for (a)$8$-mer LTHS chains, and (b)$20$-mer LTHS
chains.}
\end{figure}
\begin{figure}[!h]
\caption{Order parameter $S_{2}$ vs. volume fraction $\eta$ for LTHS
$7$-mers, comparing present theory  with Monte Carlo data[20].}
\end{figure}
\begin{figure}[!h]
\caption{Reduced pressure for LTHS $7$-mers as a function of volume
fraction $\eta$, comparing present theory with Monte Carlo data and the 
modified Vega and Lago theory from Ref.[20].}
\end{figure}
\begin{figure}[!h]
\caption{Comparison of volume fractions of the isotropic and nematic
phases at coexistence for LTHS $n$-mers, as a function of the number of
monomers $n$, between the present theory and Monte Carlo simulation data.}
\end{figure}
\begin{figure}[!h]
\caption{Comparison of the reduced pressure between this work, the Mehta and
Honnell GFD theory and the modified TPT in Ref.[8], and Monte Carlo
simulations[37] for (a)LFHS $6$-mers with bondlength to diameter
ratio $l^{*} = 0.5$, (b)LFHS $8$-mers, $l^{*} = 0.5$, and (c)LFHS
$8$-mers, $l^{*} = 0.6$.}
\end{figure}
\begin{figure}[!h]
\caption{Volume fractions of the isotropic and nematic phases at
coexistence as a function of number of monomers $n$ for LFHS chains of
constant length with $L/d = 19$.  The spherocylinder limits are given by
the Lee theory[23] and by the Monte Carlo simulation results of 
Bolhuis and Frenkel[38].}
\end{figure}
\begin{figure}[!h]
\caption{Diagrammatic representation of the excluded volume for $n = 2$.  
The slice is taken through the $z = 0$ plane, where the radius of each
circle is the monomer diameter, $d$.}
\end{figure}
\begin{figure}[!h]
\caption{Second case for two-body overlap, when  $0 < \theta_{12} <
\pi/3$.}
\end{figure}

\pagebreak


\begin{references}

\bibitem{1}  M.S.~Wertheim, J. Chem. Phys. \textbf{87,} 7323 (1987).

\bibitem{2a}  E.~Kierlik and M.L.~Rosinberg, J. Chem. Phys. \textbf{97,}
	9222 (1992).

\bibitem{2b}  E.~Kierlik and M.L.~Rosinberg, J. Chem. Phys. \textbf{99,}
	3950 (1993).

\bibitem{3}  J.~Chang and S.I.~Sandler, Chem. Eng. Sci. \textbf{49,}
	2777 (1994).

\bibitem{4}  Y.~Zhou, C.K.~Hall and G.~Stell, J. Chem. Phys.
	\textbf{103,} 2688 (1995).

\bibitem{5}  K.G.~Honnell and C.K.~Hall, J. Chem. Phys. \textbf{90,}
	1841 (1989).

\bibitem{6}  L.A.~Costa, Y.~Zhou, C.K.~Hall and S.~Carra,
	J. Chem. Phys. \textbf{102,} 6212 (1995).

\bibitem{7}  S.D.~Mehta and K.G.~Honnell, J. Phys. Chem. \textbf{100,}
	10408 (1996).

\bibitem{bn}  T.~Boublik and I.~Nezbeda, Chem. Phys. Lett. \textbf{46,}
	315 (1977);\\ T.~Boublik, Mol. Phys. \textbf{44,} 1369 (1981).

\bibitem{8}  T.~Boublik, Mol. Phys. \textbf{68,} 191 (1989).

\bibitem{17b}  A.~Yethiraj and H.~Fynewever, Mol. Phys. \textbf{93,}
	693 (1998).

\bibitem{10a}  J.G.~Curro and K.S.~Schweizer, Macromolecules. \textbf{20,}
	1928 (1987).

\bibitem{10b}  K.S.~Schweizer and J.G.~Curro, Macromolecules. \textbf{21,}
	3070,3082 (1988).

\bibitem{11}  Y.C.~Chiew, Mol. Phys. \textbf{70,} 129 (1990).

\bibitem{12}  A.~Chamoux and A.~Perera, Mol. Phys. \textbf{93,}
	649 (1998).

\bibitem{13}  S.~Sen, J.M.~Cohen, J.D.~McCoy and J.G.~Curro, 
	J. Chem. Phys. \textbf{101,} 9010 (1994).

\bibitem{14a}  C.E.~Woodward and A.~Yethiraj, J. Chem. Phys. \textbf{100,}
	3181 (1994).

\bibitem{14b}  A.~Yethiraj and C.E.~Woodward, J. Chem. Phys. \textbf{102,}
	5499 (1995).

\bibitem{15}  C.~Vega and S.~Lago, J. Chem. Phys. \textbf{100,} 6727
	(1994).

\bibitem{16}  D.C.~Williamson and G.~Jackson, J. Chem. Phys.
	\textbf{108,} 10294 (1998).

\bibitem{17a}  H.~Fynewever and A.~Yethiraj, J. Chem. Phys. \textbf{108,}
	1636 (1998).

\bibitem{18}  J.D.~Parsons, Phys. Rev. A. \textbf{19,} 1225 (1979).

\bibitem{19}  S.D.~Lee, J. Chem. Phys. \textbf{87,} 4972 (1987).

\bibitem{20}  L.~Onsager, Ann. NY Acad. Sci. \textbf{51,} 627 (1949).

\bibitem{21}  D.C.~Williamson and G.~Jackson, Mol. Phys. \textbf{86,}
	819 (1995).

\bibitem{22}  G.~Stell, in \textit{The Equilibrium Theory of Classical
        Fluids.} H.L. Frisch and J.L. Lebowitz, eds. (Benjamin, New York,
        1964).

\bibitem{23a}  D.~Chandler and L.R.~Pratt, J. Chem. Phys. \textbf{65,}
	2925 (1976).

\bibitem{23b}  L.R.~Pratt and D.~Chandler, J. Chem. Phys. \textbf{66,}
	147 (1977).

\bibitem{30}  P.~Padilla and E.~Velasco, J. Chem. Phys. \textbf{106,}
	10299 (1997).

\bibitem{31}  A.M.~Somoza and P.~Tarazona, J. Chem. Phys. \textbf{91,}
	517 (1989).

\bibitem{26}  A.~Yethiraj, J.G.~Curro, K.S.~Schweizer and J.D.~McCoy,
	J. Chem. Phys. \textbf{98,} 1635 (1993).

\bibitem{33}  Expressions for the molecular and reference volumes, $v_n$,
	$v_1$ and $v_2$, are: $
         v_n = \frac{\pi d^3}{6}\left[1 + (n - 1)\left(\frac{3}{2}l^* -
                \frac{1}{2}(l^*)^3\right)\right]$, $
         v_1 = \frac{\pi}{6}d_M^3$ and $
         v_2 = \frac{\pi d_D^3}{6}\left[1 + \frac{3}{2}l^* -
                \frac{1}{2}(l^*)^3\right]$.

\bibitem{27}  K.M.~Jaffer, M.Sc. Thesis. University of Guelph, (1999).

\bibitem{28}  N.F.~Carnahan and K.E.~Starling, J. Chem. Phys.
	\textbf{51,} 635 (1969).

\bibitem{29}  D.J.~Tildesley and W.B.~Streett, Mol. Phys. \textbf{41,}
	85 (1980).

\bibitem{34}  W.G.~Chapman, G.~Jackson and K.~Gubbins, Mol. Phys.
	\textbf{65,} 1057 (1988).

\bibitem{32}  M.~Whittle and A.J.~Masters, Mol. Phys. \textbf{72,}
	247 (1991).

\bibitem{bf}  P.~Bolhuis and D.~Frenkel, J. Chem. Phys. \textbf{106,}
        666 (1997).
\end{references}
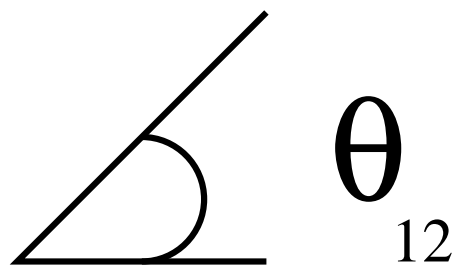
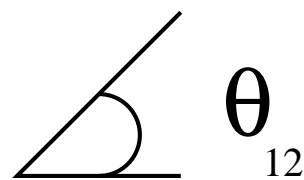
\end{document}